\documentstyle[manuscript,aps]{revtex}
\input psfig

\begin{document}
\date{July, 5. 1996}
\title{Substitution effects on spin fluctuations \\
in the spin-Peierls compound CuGeO$_3$}
\author{\bf M. Fischer$^a$, P. Lemmens$^a$, G. G\"untherodt$^a$, \\
M. Weiden$^b$, W. Richter$^b$, C. Geibel$^b$, F. Steglich$^b$} 
\address{$^a$ 2. Physikalisches Institut, RWTH-Aachen, 
52056 Aachen, Germany\\
$^b$ Physikalisches Institut, TH Darmstadt, 
64289 Darmstadt, Germany}   
\maketitle
\begin{abstract}
Using Raman scattering we studied the effect of substitutions 
on 1D spin fluctuations in CuGeO$_3$
observed as a spinon continuum in frustration induced exchange scattering. 
For temperatures below the spin-Peierls transition (T$_{SP}$=14K)
the intensity of this continuum at 120-500 cm$^{-1}$ is exponentially 
suppressed and transferred into a 3D two-magnon density of states. 
Besides a spin-Peierls gap-induced mode at 30 cm$^{-1}$ and additional modes 
at 105 and 370 cm$^{-1}$ are observed. 
Substitution of Zn on the Cu-site and Si on the Ge-site of CuGeO$_3$ 
quenches easily the spin-Peierls state. 
Consequently a suppression of the spin-Peierls gap observable
below T$_{SP}$=14K as well as a change of
the temperature dependence of the spinon continuum are observed.
These effects are discussed in the context of a dimensional 
crossover of this compound below T$_{SP}$ and strong spin-lattice interaction.
\end{abstract}
Keywords: Spin-Peierls transition, Raman scattering, spin fluctuations, spinon continuum\\
Author for correspondance:
P. Lemmens, 2. Phys. Inst. RWTH Aachen, Templergraben 55, 52056 Aachen\\
Tel.: 241-807113, Fax.: 241-8888306, Email: lemmens@physik.rwth-aachen.de \\
\section{INTRODUCTION}
Quantum 1D spin systems show a variety of instabilities. Of particular interest is the
spin-Peierls (SP) transition due to finite magnetoelastic couplings, 
leading to the opening of a gap in the spinon excitation continuum 
\cite{Mueller,KCuF3}.
The discovery \cite{Hase} of the spin-Peierls transition below T$_{SP}$=14 K 
in an inorganic compound, CuGeO$_3$, has attracted widespread attention 
\cite{Nishi,Raman,Lemmens}. 
This compound consists of chains of spin-1/2 Cu$^{2+}$ ions coupled by 
antiferromagnetic superexchange through oxygen orbitals \cite{Hase,Nishi}. 
The exchange along these chains can be modeled by a 1D-Hamiltonian 
including the competing nearest and next-nearest neighbor interactions 
\cite{Castilla}. 
As shown recently by exact diagonalization and comparison with 
experimental data the Raman process for scattering in the 
A$_{1g}$-contribution with polarization parallel to the c-axis and the Cu-chains 
is described by frustrated Heisenberg exchange scattering \cite{PRL} if for 
temperatures below $T_{SP}$ a dimerization 
of the individual exchange coupling constants is taken into account. 
In this way the observed scattering continuum 
from 120-500 cm$^{-1}$ (above and below T$_{SP}$) and an additional excitation 
at 30 cm$^{-1}$ (below T$_{SP}$) related to the spin-Peierls gap ($\approx$2$\Delta$) 
can be modeled by 1D-spinon excitations. 
All other scattering contributions should be negligible. 
However, the observed quasi-3D two-magnon density of states in Raman scattering 
and the related magnon dispersion in neutron scattering both observed 
below T$_{SP}$ \cite{Nishi} remain to be explained. 
This dimensional crossover from 1D to 3D upon cooling 
and the observation of a strong sensitivity of 
the spin-Peierls state to any substitution with a 
subsequent magnetically ordered state at lower temperatures (T$<4K$) 
are pointing to an interplay of the 
the spin-Peierls order parameter with 3D magnetic interactions \cite{PRL}. 
This motivated our Raman study on differently 
substituted samples of CuGeO$_3$ \cite{Weiden}.
\section{EXPERIMENTAL}
We have performed Raman scattering experiments on Zn- and Si-substituted 
single crystals in quasi-back\-scat\-tering 
geometry with the polarization of incident and scattered 
light parallel to the c-axis and the Cu-O chains. In other 
scattering geometries no magnetic contribution was observed.
Here we present results on (Cu$_{1-x}$,Zn$_x$)GeO$_3$ and Cu(Ge$_{1-y}$,Si$_y$)O$_3$.
Details of the experiment will be published elsewhere \cite{Lemmens_long}. 
\section{RESULTS AND DISCUSSION}
With Zn-substitution on the Cu-site the spin chains are effectively cut.
The formation of spin singlets is locally inhibited resulting
in a strong depression of the spin-Peierls-transition.
The effect of Si-substitution on the Ge-site is mediated via the
bridging oxygen leading to a similar local disruption of the exchange path. Due
to a different ionic radius of Si compared to Ge lattice strains are induced.

Depending on the substitution more or less appreciable shifts of the Raman-active 
phonons were observed. In good agreement with x-ray scattering the biggest 
lattice strains are induced by Si-substitution leading to linear shifts of several
phonons in dependence of the substitution level. 
Zn-substitution on the Cu-site leads for $x=0.02-0.03$ to a minimum  
of the phonon frequencies for the phonons observed at 186, 227, 330 and 430 cm$^{-1}$. 
This suggests a spin-phonon coupling of these modes,  
may be due to their energy close to the spinon excitation spectrum \cite{Weiden}.
Additionally to the static properties of the lattice the spinon 
excitations are changed, too.
In Fig. 1 Raman spectra of (Cu$_{1-x}$,Zn$_x$)GeO$_3$ at T=6.2K are shown for different x.
With increasing Zn-substitution the gap-related mode at 30 cm$^{-1}$
(observed below T$_{SP}$) is suppressed in intensity. This coincides with
the vanishing of the spin-Peierls gap. A similar behavior is observed for Si-substitution.
A mode at 107 cm$^{-1}$ attributed to a spin-Peierls active phonon
with a Fano-lineshape vanishes for a substitution level higher than $x=0.02$.
A broad maximum at 227 cm$^{-1}$ which can be modelled 
by a 3D two-magnon-density of states is suppressed but still
survives even for the highest substitution of $x=0.06$.
For this Zn-concentration magnetic susceptibility measurements
show no Peierls transition anymore. The mode at 370 cm$^{-1}$ attributed
to a spin-Peierls active phonon is depressed but survives, too. So the spin-Peierls active
modes react in a different way on substitution.
The spinon continuum observed as a very broad maximum in the frequency
range 120-500 cm$^{-1}$ is observed both in clean and substituted samples.
We determined its integrated intensity by subtracting phonons,
two-magnon-contributions and a linear background using an integration
in the frequency range 120-500 cm$^{-1}$. The data was subsequently normalized to the
intensity of a phonon at 594 cm$^{-1}$ to facilitate the comparison
between different x. \\
As shown in Fig. 2 the normalized integrated intensity as function of temperature
shows a rise toward lower temperature, a plateau and a sharp drop
below $T_{SP}$ in the clean sample, $x=0$. This drop marks the opening
of the spin-Peierls gap freezing out spinon excitations. 
No drop is observed for sufficiently large Zn- ($x > 0.03$) or Si-substitution ($y > 0.02$).
The survival of the continuum, independent of substitution, clearly stresses
the short-range character of the excitations along the Cu-chain. It is remarkable
that the normalised integrated intensity for Zn-substitution is at its maximum
comparable to the clean sample. The Si-substituted samples show different behavior.
Here this value is lower than for the pure case. This remains to be explained.

Substitution on the Ge- or Cu-site does not change the value of the superexchange
$J$ as one neither observes a shift in
the maximum of the susceptibility \cite{Weiden} nor in the position of the
spinon continuum. In general it is quite striking that only a small amount of doping is required
to suppress the spin-Peierls transition completely independent of the specific mechanism. 
The important point seems to be that the system is disturbed
by substitution. This is an indication for a moderate energy gain 
changing from the uniform to the dimerized phase. Frustration taking
into account next-nearest neighbor interactions may explain a larger
influence of local defects on the ground state.

CuGeO$_{3}$ is considered to be a quasi-1D frustrated Heisenberg-antiferromagnet.
Also the spin-Peierls transition is an effect clearly attributed to 1D systems.
However, not all spin-Peierls active modes can be considered as 1D.
Both the 2-magnon density of states below 227 cm$^{-1}$ and the additional
phonon modes at 370 cm$^{-1}$ and 820 cm$^{-1}$ have a 3D character,
while the gap related 30 cm$^{-1}$ mode and the spinon continuum are attributed
to 1D behavior. So the arise of the spin-Peierls transition in this compound
is affected by an interplay between the 1D spin chains, the 3D
phonon system and a 3D behavior of the spin system upon cooling below T$_{SP}$. 
Substitution offers the chance to observe 
this interplay as it  suppresses the spin-Peierls state and as well induces a 3 dimensional
Neel state for higher levels of substitution. Therefore it is necessary to further
investigate and understand
the temperature dependence of the spin-Peierls active modes for substitutions.
Results of our studies will be published elsewhere \cite{Lemmens_long}.    

\section{CONCLUSIONS}
Using Raman scattering the spinon excitations observed 
as a gap-related signal and a spinon continuum are studied.
They respond differently to changes of the temperature 
and to substitutions. While the gap-related signal showing up
below T$_{SP}$ strongly decreases in intensity upon substitution,
the spinon continuum above T$_{SP}$ shows no changes. However,
the drastic decrease of the integrated continuum below T$_{SP}$ is
suppressed in the substituted samples due to the suppression of the dimerized 
phase. This effect does not really depend on the kind of defect giving rise
to the supposition that there is an universal phase diagram for substituted CuGeO$_{3}$
samples.

Acknowledgement: This work was supported DFG, through SFB 341 and SFB 252 and by BMBF 13N6586/8.
%
%

\newpage
{\bf Figure Caption}\\

Fig 1.: Scattering intensity for Cu$_{1-x}$Zn$_x$GeO$_3$ 
with x=0, 0.018 and 0.06 at T=6.2K. The spectra for x=0 and x=0.018 are
shifted by 200 and 100 a.u. respectively.\\

Fig 2.: Temperature dependence of the normalised integrated 
intensity of the spinon continuum for Cu$_{1-x}$Zn$_x$GeO$_3$ 
with x=0, 0.018, 0.06 and Cu(Ge$_{1-y}$,Si$_y$)O$_3$ for y=0.022.
The curve for x=0.06 is shifted by 2900 a.u.

\newpage
\begin{figure}
\centerline{\psfig{file=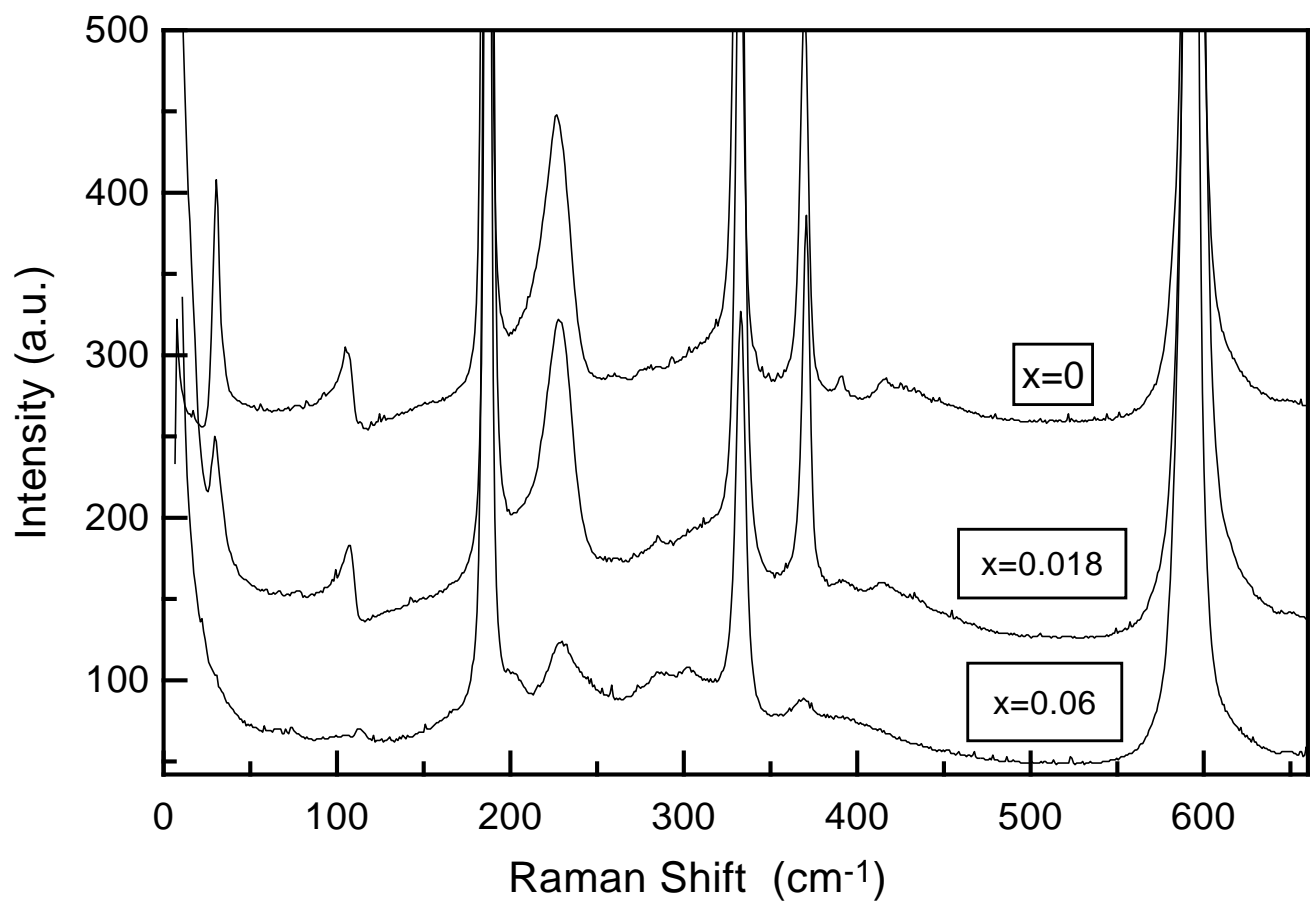,height=14cm,rheight=-10cm}}
\caption{Scattering intensity for Cu$_{1-x}$Zn$_x$GeO$_3$ 
with x=0, 0.018 and 0.06 at T=6.2K. The spectra for x=0 and x=0.018 are
shifted by 200 and 100 a.u. respectively.}
\label{f1}
\end{figure}
\newpage
\begin{figure}
\centerline{\psfig{file=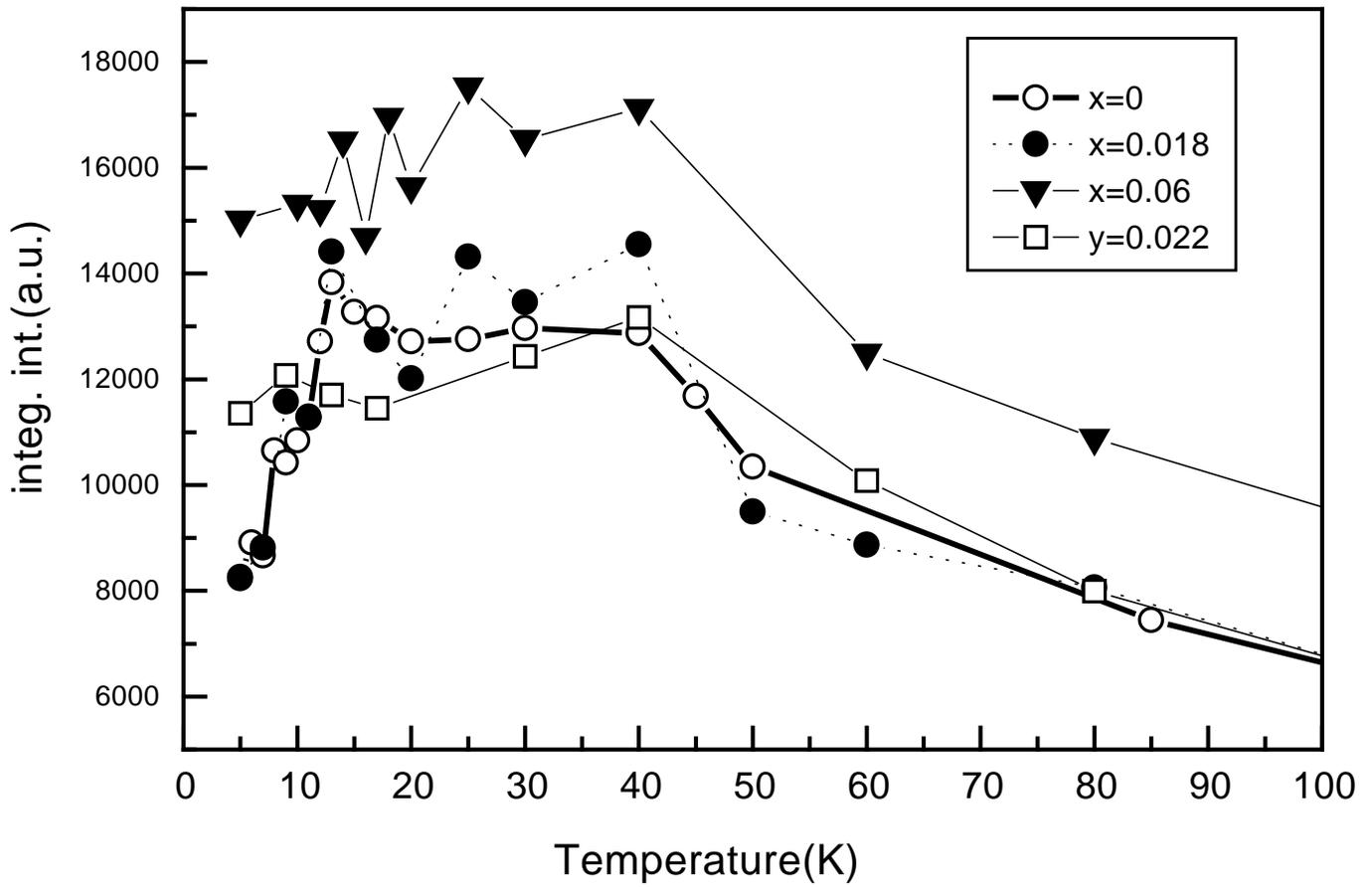,height=14cm,rheight=-10cm}}
\caption{Temperature dependence of the normalised integrated 
intensity of the spinon continuum for Cu$_{1-x}$Zn$_x$GeO$_3$ 
with x=0, 0.018, 0.06 and Cu(Ge$_{1-y}$,Si$_y$)O$_3$ for y=0.022.
The curve for x=0.06 is shifted by 2900 a.u.}
\label{f1}
\end{figure}

\end{document}